\documentclass[numsec,webpdf,modern,large,namedate]{oup-authoring-template}

\theoremstyle{thmstyleone}%
\theoremstyle{thmstyletwo}%
\theoremstyle{thmstylethree}%

\usepackage{hyperref}
\hypersetup{colorlinks=true, citecolor=blue, urlcolor=blue, linkcolor=blue}
\usepackage{graphicx}
\usepackage{multirow}
\usepackage{setspace}
\usepackage{bm}
\usepackage{amsmath,amsfonts}
\usepackage{subfigure} 

\newcommand{\ie}{\textit{i}.\textit{e}., }
\newcommand{\eg}{\textit{e}.\textit{g}. }

\begin{document}

\journaltitle{Bioinformatics}
\DOI{https://doi.org/xxx}
\copyrightyear{xxxx}
\appnotes{\bf ISMB 2024}

\firstpage{1}

\title{RiboDiffusion: Tertiary Structure-based RNA Inverse Folding with Generative Diffusion Models}

\author[1,2,$\dagger$]{Han Huang}
\author[1,3,$\dagger$]{Ziqian Lin}
\author[1]{Dongchen He}
\author[1]{Liang Hong}
\author[1,$\ast$]{Yu Li}

\authormark{Huang, H., et al.}

\address[1]{Department of Computer Science and Engineering, CUHK, Hong Kong SAR, China}
\address[2]{School of Computer Science and Engineering, Beihang University, Beijing, China}
\address[3]{Nanjing University, Nanjing, China}

\corresp[$\ast$]{Corresponding author.}
\corresp[$\dagger$]{Equal contribution.}

\abstract{RNA design shows growing applications in synthetic biology and therapeutics, driven by the crucial role of RNA in various biological processes. 
A fundamental challenge is to find functional RNA sequences that satisfy given structural constraints, known as the inverse folding problem.
Computational approaches have emerged to address this problem based on secondary structures. 
However, designing RNA sequences directly from 3D structures is still challenging, due to the scarcity of data, the non-unique structure-sequence mapping, and the flexibility of RNA conformation.
In this study, we propose RiboDiffusion, a generative diffusion model for RNA inverse folding that can learn the conditional distribution of RNA sequences given 3D backbone structures.
Our model consists of a graph neural network-based structure module and a Transformer-based sequence module, which iteratively transforms random sequences into desired sequences.
By tuning the sampling weight, our model allows for a trade-off between sequence recovery and diversity to explore more candidates.
We split test sets based on RNA clustering with different cut-offs for sequence or structure similarity.
Our model outperforms baselines in sequence recovery, with an average relative improvement of $11\%$ for sequence similarity splits and $16\%$ for structure similarity splits.
Moreover, RiboDiffusion performs consistently well across various RNA length categories and RNA types. 
We also apply in-silico folding to validate whether the generated sequences can fold into the given 3D RNA backbones. Our method could be a powerful tool for RNA design that explores the vast sequence space and finds novel solutions to 3D structural constraints.}


\maketitle

\section{Introduction}


The design of RNA molecules is an emerging tool in synthetic biology \citep{chappell2015renaissance, mckeague2016opportunities} and therapeutics \citep{zhu2022rna}, enabling the engineering of specific functions in various biological processes. 
There have been various explorations into RNA-based biotechnology,
such as translational RNA regulators for gene expression \citep{lagana2015computational, chappell2017computational}, aptamers for diagnostic or therapeutic applications \citep{espah2016automated, findeiss2017design}, and catalysis by ribozymes \citep{dotu2014complete, park2019catalytic}. 
While the tertiary structure determines how RNA molecules function, one fundamental challenge in RNA design is to create functional RNA sequences that can fold into the desired structure, also known as the inverse RNA folding problem \citep{hofacker1994fast}.


Most early computational methods for inverse RNA folding focus on folding into RNA secondary structures \citep{churkin2018design}. 
Some programs use efficient local search strategies to optimize a single seed sequence for the desired folding properties, guided by the energy function \citep{hofacker1994fast, andronescu2004new,  busch2006info, garcia2013rnaifold}. 
Others attempt to solve the problem globally by modeling the sequence distribution or directly manipulating diverse candidates \citep{taneda2010modena, kleinkauf2015antarna, yang2017rna, runge2019learning}.    
However, without considering 3D structures of RNA, these methods cannot meet accurate functional structure constraints, since RNA secondary structures only partially determine their tertiary structures \citep{vicens2022thoughts}.
The pioneering work \citep{yesselman2015rna} applies a physically-based approach to optimize RNA sequences and match the fixed backbones, but it is still constrained by the local design strategy and computational efficiency.

Recent advances in deep learning and the accumulation of biomolecular structural data have enabled computational methods to model mapping between sequences and 3D structures with extraordinary performance, as demonstrated by remarkable results in protein 3D structure prediction \citep{jumper2021highly, lin2023evolutionary} and inverse design \citep{dauparas2022robust}.
Inspired by this, the development of geometric learning methods on RNA structures has received increasing research interest.
On the one hand, many studies have explored RNA tertiary structure prediction using machine learning models with limited data \citep{shen2022e2efold, baek2022accurate, li2023integrating}.
On the other hand, although deep learning has a promising potential to narrow down the immense sequence space for inverse folding, developing an appropriate model for RNA inverse folding remains an open problem, as it requires capturing the geometric features of flexible RNA conformations, handling the non-unique mappings between structures and sequences, and providing alternative options for different design preferences.

In this study, we introduce RiboDiffusion, a generative diffusion model for RNA inverse folding based on tertiary structures.
We formulate the RNA inverse folding problem as learning the sequence distribution conditioned on fixed backbone structures, using a generative diffusion model \citep{yang2022diffusion}.
Unlike previous methods that predict the most probable sequence for a given backbone \citep{ingraham2019generative, jing2020learning, gao2022pifold, joshi2023multi}, our method captures multiple mappings from 3D structures to sequences through distribution learning.
With a generative denoising process for sampling, our model iteratively transforms random initial RNA sequences into desired candidates under tertiary structure conditioning.
This global iterative generation distinguishes our model from autoregressive models and local updating methods, enabling it to better search for sequences that satisfy global geometric constraints.   
We parameterize the diffusion model with a cascade of a structure module and a sequence module, to capture the mutual dependencies between sequence and structure.
The structure module, based on graph neural networks, extracts SE(3)-invariant geometrical features from 3D fixed RNA backbones, while the sequence module, based on Transformer-liked layers, captures the internal correlations of RNA primary structures.
To train the model, we randomly drop the structural module to learn both the conditional and unconditional RNA sequence distribution.
We also mix the conditional and unconditional distributions in the sampling procedures, to balance sequence recovery and diversity for more candidates.

We use RNA tertiary structures from PDB database \citep{bank1971protein} to construct the benchmark dataset and augment it with predicted structures from the RNA structure prediction model \citep{shen2022e2efold}.  
We split test sets based on RNA clustering using different sequence or structure similarity cutoffs.
Our model achieves an $11\%$ higher recovery rate than the machine learning baselines for benchmarks based on sequence similarity, and $16\%$ higher for benchmarks based on structure similarity. 
RiboDiffusion also performs consistently well across different RNA lengths and types. Further analysis reveals its great performance for cross-family and in-silico folding. 
Our method could be a powerful tool for RNA design, exploring a wide sequence space and finding novel solutions to 3D structural constraints.


\section{Methodology}

\begin{figure*}[t]
\centering

\includegraphics[width=0.98\textwidth]{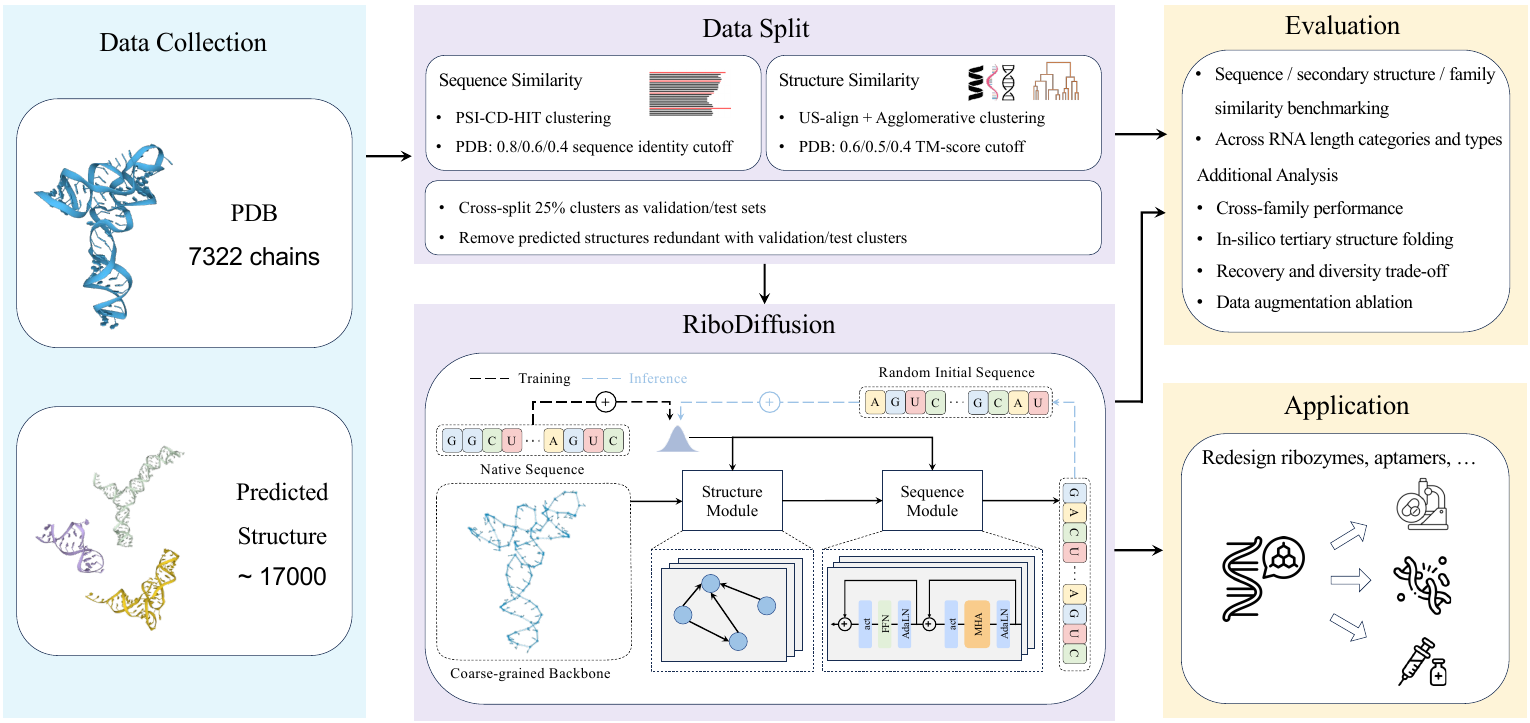}

\caption{Overview of RiboDiffusion for tertiary structure-based RNA inverse folding.
We construct a dataset with experimentally determined RNA structures from PDB, supplemented with additional structures predicted by an RNA structure prediction model.
We cluster RNA with different cut-offs for sequence or structure similarity and make cross-split to evaluate models.
RiboDiffusion trains a neural network with a structure module and a sequence module to recover the original sequence from a noisy sequence and a coarse-grained RNA backbone extracted from the tertiary structure.
RiboDiffusion then uses the trained network to iteratively refine random initial sequences until they match the target structure.
We present a comprehensive evaluation and analysis of the proposed method.
}
\label{fig:ribo_train}
\end{figure*}

This section will explain RiboDiffusion in detail - a deep generative model for RNA inverse folding based on fixed 3D backbones. 
The overview is shown in \autoref{fig:ribo_train}. We will first introduce the preliminaries of diffusion models and our formulations of the RNA inverse folding problem. We will then describe the design of neural networks to parameterize the diffusion model and explain the sequence sampling procedures.

\subsection{Preliminary and Formulation}

\subsubsection{Diffusion Model}
As a powerful genre of generative models, diffusion models \citep{sohl2015deep} have been successfully applied to the distribution learning of diverse data, including images \citep{ho2020denoising, song2021score}, graphs \citep{huangGraphgdp22, Huang_CDGS_2023}, and molecular geometry \citep{watson2023novo, huang2023learning}.
As the first step of setting up the diffusion model, a forward diffusion process is constructed to perturb data with a sequence of noise. This converts the data distribution to a known prior distribution.
With random variables $\mathrm{x}_0 \in \mathbb{R}^{d}$ and a forward process $\{\mathrm{x}_t\}_{t \in [0,T]}$, a Gaussian transition kernel is set as
\begin{equation}
q_{0t}(\mathrm{x}_t|\mathrm{x}_0) = \mathcal{N}(\mathrm{x}_t | \alpha_t \mathrm{x}_0, {\sigma}^2_t \bm{I}) \ ,
\label{eq:q0t}
\end{equation}
where $\alpha_t, \sigma_t \in \mathbb{R}^{+} $ are time-dependent differentiable functions that are usually chosen to ensure a strictly decreasing signal-to-noise ratio (SNR) $\alpha^2_t / \sigma^2_t$ and the final distribution $q_T(\mathrm{x}_T) \approx \mathcal{N}(\bm{0}, \bm{I})$ \citep{VDM2021}. 
Diffusion models can generate new samples starting from the prior distribution, after learning to reverse the forward process. 
Such the reverse-time denoising process from time $T$ to time $0$ can be described by a stochastic differential equation (SDE) \citep{yang2022diffusion} as
\begin{equation}
    \mathrm{d} \mathrm{x}_t = [f(t) \mathrm{x}_t - g^2(t) \nabla_{\mathrm{x}}\log p_t (\mathrm{x}_t)] \mathrm{d}_t 
    + g(t) \mathrm{d} \bar{\bm{w}}_t \ ,
\label{eq:reSDE_ori}
\end{equation}
where $\nabla_{x}\log p_t (x_t)$ is the so-called score function and $\bar{\bm{w}}_t$ is the standard reverse-time Wiener process. While $f(t)=\frac{\mathrm{d} \log \alpha_t} {\mathrm{d}t} $ is the drift coefficient of SDEs,  $g^2(t)= \frac{\mathrm{d}\sigma^2_t}{\mathrm{d}t} - 2\frac{\mathrm{d} \log \alpha_t} {\mathrm{d}t}\sigma^2_t$ is the diffusion coefficient \citep{VDM2021}.
Deep neural networks are used to parameterize the score function variants in two similar forms, \ie the noise prediction model $\bm{\epsilon}_{\bm{\theta}}(\mathrm{x}_t, t)$ and the data prediction model $\bm{d}_{\bm{\theta}}(\mathrm{x}_t, t)$. 
In this study, we focus on the parameterization of the widely used data prediction model to directly predict the original data $\mathrm{x}_0$ from $\mathrm{x}_t$.

\subsubsection{RNA Inverse Folding}
Inverse folding aims to explore sequences that can fold into a predefined structure, which is specified here as the fixed sugar-phosphate backbone of an RNA tertiary structure. 
For an RNA molecule with $N$ nucleotides consisting of four different types A (Adenine), U (Uracil), C (Cytosine), and G (Guanine), its sequence can be defined as
$\bm{S} \in \{\mathrm{A}, \mathrm{U}, \mathrm{C}, \mathrm{G}\}^N$.
Among the backbone atoms, we choose one three-atom coarse-grained representation including the atom coordinates of C4', C1', N1 (pyrimidine) or N9 (purine) for every nucleotide.
The simplified backbone structure can be denoted as $ \bm{X} \in \mathbb{R}^{3N \times 3} $.
Note that there are various alternative schemes for coarse-graining RNA 3D backbones, including using more atoms to obtain precise representations \citep{dawson2016coarse}. We explore a concise representation with regular structural patterns \citep{shen2022e2efold}.

Formally, we consider the RNA inverse folding problem as modeling the conditional distribution $p(\bm{S} | \bm{X})$, \ie the sequence distribution conditioned on RNA backbone structures.
We establish a diffusion model to learn the conditional sequence distribution.
To take advantage of the convenience of defining diffusion models in continuous data spaces \citep{chen2022analog, dieleman2022continuous}, discrete nucleotide types in the sequence are represented by one-hot encoding and continuousized in the real number space as $\bm{S} \in \mathbb{R}^{4N}$.
The continuous-time forward diffusion process in the sequence space $\mathbb{R}^{4N}$ can be described by the forward SDE with $t \in [0,T]$ as
$\mathrm{d} \bm{S}_t = f(t) \bm{S}_t \mathrm{d}t + g(t) \mathrm{d}\bm{w}$.
Under this forward SDE, the original sequence at time $t=0$ is gradually corrupted by adding Gaussian noise. 
With the linear Gaussian transition kernel derived from the forward SDE in Eq. (\ref{eq:q0t}) \citep{yang2022diffusion}, we can conveniently sample $\bm{S}_t = \alpha_t + \sigma_t \epsilon_{\bm{S}}$ at any time $t$ for training, where $\epsilon_{\bm{S}}$ is Gaussian noise in the sequence space.
For the generative denoising process, the corresponding reverse-time SDE from time T to 0 can be derived from Eq. (\ref{eq:reSDE_ori}) as
\begin{equation}
    \mathrm{d} \bm{S}_t = [f(t) - g^2(t) \nabla_{\bm{S}}\log p_t (\bm{S}_t | \bm{X})] \mathrm{d}t 
    + g(t) \mathrm{d}(\bar{\bm{w}}_{t}) \ ,
\label{eq:reverse_SDE}
\end{equation}
where $p_t (\bm{S}_t | \bm{X})$ is the marginal distribution of sequences given $\bm{X}$, and the score function $\nabla_{\bm{S}}\log p_t (\bm{S}_t | \bm{X})$ represents the gradient field of the logarithmic marginal distribution.

Once the score function is parameterized, we can numerically solve this reverse SDE to convert random samples from the prior distribution $\mathcal{N}(\bm{0}, \bm{I})$ into the desired sequences.
We establish a data prediction model to achieve the score function parameterization, learning to reverse the forward diffusion process. Specifically, we feed the noised sequence data $\bm{S}_t$, the log signal-to-noise ratio $\lambda_t = \log(\alpha^2_t / \sigma^2_t)$, and the conditioning RNA backbone structures $\bm{X}$ to the data prediction model $\bm{d_\theta}(\bm{S}_t, \lambda_t, \bm{X})$.
We optimize the data prediction model with a simple weighted squared error objective function:
\begin{equation}
\begin{aligned}
    \min_{\bm{\theta}} \mathbb{E}_t\{ \sqrt{\frac{\alpha_t}{\sigma_t}} \mathbb{E}_{\bm{S}_0, \bm{X}} \mathbb{E}_{\bm{S}_t|\bm{S}_0} 
    ||\bm{d_\theta}(\bm{S}_t, \lambda_t, \bm{X}) - \bm{S}_0 ||_2^2 \} \ ,
\end{aligned}
\end{equation}
which can be considered as optimizing a weighted variational lower bound on the data log-likelihood or a form of denoising score matching \citep{ho2020denoising, song2021score, VDM2021}.

\subsection{Model Architecture}

The architecture design of the data prediction model largely determines the diffusion learning quality of the diffusion model.
We propose a two-module model to predict the original nucleotide types: a structure module to capture geometric features and a sequence module to capture intra-sequential correlation.

\subsubsection{Structure Module}
Geometric deep learning models aim to extract equivariant or invariant features from 3D data and achieve impressive performance in the protein inverse folding task \citep{ingraham2019generative, jing2020learning, gao2022pifold}.
Our structure module is constructed based on the GVP-GNN architecture \citep{jing2020learning} and adapted for RNA backbone structures.

The fixed RNA backbone is first represented as a geometric graph $\mathcal{G} = (\mathcal{V}, \mathcal{E})$ where each node $v_i \in \mathcal{V}$ corresponds to a nucleotide and connects to its top-$k$ nearest neighbors according to the distance of C1' atoms.
The scalar and vector features are extracted from 3D coordinates as node and edge attributes in graphs, which describe the local geometry of nucleotides and their relative geometry.
Specifically, the scalar node features in nucleotides are obtained from dihedral angles, while the vector node features consist of forward and reverse vectors of sequential C1' atoms, as well as the local orientation vectors of C1' to C4' and N1/N9.
The initial embedding of each edge consists of its connected C1' atom's direction vector, Gaussian radial basis encoding for their Euclidean distance, and sinusoidal position encoding \citep{vaswani2017attention} of the relative distance in the sequence.  
In addition to geometry information, we also append the corrupted one-hot encoding of nucleotide types $\bm{S}_t$ as the node scalar features.
Furthermore, inspired by the widely used self-conditioning technique in diffusion models \citep{chen2022analog, watson2023novo, huang2023learning}, the previously predicted sequence output, denoted as $\tilde{\bm{S}_0}$, is also considered as node embeddings to enhance the utilization of model capacity.
To update the node embeddings, the nucleotide graph employs a standard message passing technique \citep{gilmer2017neural}. This involves combining the neighboring nodes and edges through GVP layers, where scalar and vector features interact via gating to create messages. The resulting messages are then transmitted across the graph to update scalar and vector node representations.

\begin{algorithm}[t]
\caption{RiboDiffusion Training.}
\label{alg:training}

\begin{algorithmic}[1]
    \State $t \sim \mathcal{U}(0,1], \ \bm{S}_0, \bm{X} \sim \mathrm{Training \ Set}$
    \State $\bm{S}_t \sim \mathcal{N}(\bm{S}_t | \alpha_t \bm{S}_0, \sigma_t^2 \bm{I}), \
            \lambda_t = \log(\alpha^2_t / \sigma^2_t),\ \tilde{\bm{S}}_0 \gets \bm{0}$
    \If{$\mathrm{Uniform}(0, 1.0) < 0.5 $} \Comment{Self Conditioning}
    \State $\tilde{\bm{S}_0} \gets \bm{d_{\theta}}([\bm{S}_t, \tilde{\bm{S}_0}], \lambda_t, \bm{X})$
    \State $\tilde{\bm{S}}_0 \gets \mathrm{StopGradient}(\tilde{\bm{S}_0})$
    \EndIf
    \If{$\mathrm{Uniform}(0, 1.0) < 0.4 $} \Comment{Drop Structure Condition}
    \State $\bm{X} \gets \bm{0}$ 
    \EndIf
    
    \State Minimize $\sqrt{\frac{\alpha_t}{\sigma_t}}\ [||\bm{d_{\theta}}([\bm{S}_t, \tilde{\bm{S}_0}], \lambda_t, \bm{X}) - \bm{S}_0||^2_2 ]$
\end{algorithmic}
\end{algorithm}

\begin{algorithm}[t]
\caption{RNA inverse folding via RiboDiffusion.}
\label{alg:sampling}
\textbf{Require}: time schedule $\{t_i\}_{i=0}^M$, RNA backbone coordinates $\bm{X}$

\begin{algorithmic}[1]
\State $\tilde{\bm{S}_0} \gets \bm{0}$
\State $\bm{S}_{t_0} \gets \bm{S}_{T} \sim \mathcal{N}(\bm{0}, \bm{I})$

\For{$i \gets 1$ to $M$}
    \State $t \gets t_{i-1},\  s \gets t_{i},\  \lambda_t \gets \log(\alpha^2_t / \sigma^2_t)$
    \State $\alpha_{t|s} \gets \alpha_t / \alpha_s,\ \sigma_{t|s}^2 \gets \sigma_t^2 - \alpha_{t|s}^2 \sigma_s^2$
    \State $\tilde{\bm{S}_0} \gets \bm{d_\theta}([\bm{S}_t, \tilde{\bm{S}_0}], \lambda_t, \bm{X})$
    \State $\bar{\bm{S}}_s \gets \frac{\alpha_{t|s} \sigma_s^2}{\sigma_t^2} \bm{S}_{t} + \frac{\alpha_s \sigma_{t|s}^2}{\sigma_t^2} \tilde{\bm{S}}_0$
    \State $\bm{S}_{\bm{\epsilon}} \sim \mathcal{N}(\bm{0}, \bm{I})$    
    \State $\bm{S}_s \gets \bar{\bm{S}}_s + \frac{\sigma_{t|s}\sigma_s}{\sigma_t}\bm{S}_{\bm{\epsilon}}$
\EndFor
\State \Return $\bar{\bm{S}}_{t_M}$
\end{algorithmic}
\end{algorithm}

\subsubsection{Sequence Module}
The sequential correlation in RNA primary structures is crucial for inverse folding and to obtain high-quality RNA sequences even with imprecise 3D coordinates. This concept is applicable in the inverse folding of proteins \citep{hsu2022learning, zheng2023structure}.
The sequence module takes in $f$-dimensional nucleotide-level embeddings ${\mathbf{h}}^0 \in \mathbb{R}^{N \times f}$ as tokens, which consists of SE(3)-invariant scalar node representations from the structure module and corrupted sequence data. During training, we randomly add self-conditioning sequence data similar to those of the structure module and drop structural features to model both the conditional and unconditional sequence distributions for further application. 

Our sequence module architecture is modified from the Transformer block \citep{vaswani2017attention} to inject diffusion context, log-SNR $\lambda$, or other potential conditional features (\eg RNA types) \citep{dhariwal2021diffusion, peebles2023scalable}.  
The context input $\mathbf{C}$ affects sequence tokens in the form of adaptive normalization and activation layers, which are denoted as adaLN and act functions:
\begin{equation}
\begin{aligned}
&\mathrm{adaLN}(\mathbf{h}, \mathbf{C}) = (\bm{1}+\mathrm{MLP}_1(\mathbf{C})) \cdot \mathrm{LN}(\mathbf{h}) + \mathrm{MLP}_2(\mathbf{C}), \\
&\mathrm{act}(\mathbf{h}, \mathbf{C}) = \mathrm{MLP}_3(\mathbf{C}) \cdot \mathbf{h},
\end{aligned}
\end{equation}
where $\mathrm{LN}(\cdot)$ is the layer normalization and $\mathrm{MLP}(\cdot)$ is a multilayer perception to learn shift and scale parameters.
The $l$-th Transformer block 
is defined as follows
\begin{equation}
\begin{aligned}
    & \mathbf{m}^{l} = \mathrm{MHA}(\mathrm{adaLN}(\mathbf{h}^l, \lambda_t)) \ , \\
    & {\mathbf{h}^{l+1}}' = \mathrm{act}(\mathbf{m}^{l}, \lambda_t) + \mathbf{h}^{l}, \\
    & \mathbf{h}^{l+1} = \mathrm{act}(\mathrm{FFN}(\mathrm{adaLN}({\mathbf{h}^{l+1}}', \lambda_t)), \lambda_t) + {\mathbf{h}^{l+1}}', \\
\end{aligned}
\end{equation}
where $\mathrm{MHA}(\cdot)$ is the multi-head attention layer and $\mathrm{FFN}(\cdot)$ is the Feedforward neural network \citep{vaswani2017attention}.
Finally, the sequence module output $\mathbf{h}^{L}$ is projected to nucleotide one-hot encodings via an extra MLP. 
The detailed training procedure is referred to as \autoref{alg:training}.

\subsection{Sequence Sampling}

To generate RNA sequences that are likely to fold into the given backbone, we construct a generative denoising process based on the parameterized reverse-time SDE with the optimized data prediction model $\bm{d_\theta}$, as described in Eq. (\ref{eq:reverse_SDE}).
Various numerical solvers for the SDE can be employed for sampling, such as ancestral sampling, the Euler-Maruyama method, etc.
We apply convenient ancestral sampling combined with the data prediction model and self-conditioning to generate sequences.
\autoref{alg:sampling} outlines the specific sampling procedure.
For more details on the noise schedule parameters, including $\alpha_t$ and $\sigma_t$, refer to \citep{VDM2021}.
We intuitively explain the denoising process as follows: we start by sampling noisy data from a Gaussian distribution that represents a random nucleotide sequence, and we iteratively transform this data towards the desired candidates under the condition of the given RNA 3D backbones.

Exploring novel RNA sequences that fold into well-defined 3D conformations distinct from the natural sequence is also an essential goal for RNA design, as it has the potential to introduce new functional sequences.
This task not only requires the model to generate sequences that satisfy folding constraints but also to increase diversity for subsequent screening.
During the generative denoising process, our model can balance the proportion of unconditional and conditional sequence distributions by adjusting the output of the data prediction model.
Let $w$ be the conditional scaling weight, and the data prediction model can be modified as
\begin{equation}
    \tilde{\bm{d_\theta}}(\bm{S}_t, \lambda_t, \bm{X}) = 
    w {\bm{d_\theta}}(\bm{S}_t, \lambda_t, \bm{X}) + (1-w) {\bm{d_\theta}}(\bm{S}_t, \lambda_t, \bm{0}).
\end{equation}
Setting $w=1$ is the original conditional data prediction model while decreasing $w<1$ weakens the effect of conditional information and strengthens the sequence diversity.
In this way, we achieve a trade-off between recovering the original sequence and ensuring diversity.
The distribution weighting technique is also used in diffusion models for text-to-image generation \citep{ho2022classifier, saharia2022photorealistic}.

\section{Results}

\begin{table*}[t]
\centering
\caption{
Recovery rate (\%) comparison across six different settings.
The average and standard deviation values of model performance on four random-split non-overlapping test sets are reported.
Mean recovery rates are reported for short (L$<=$50nt), medium (50nt$<$L$<=$100nt), and long (L$>$100nt) RNA.
\textit{Seq. 0.8}: sequence similarity-based split with 0.8 cluster threshold. \textit{Struct. 0.6}: structure similarity-based split with 0.6 cluster threshold.
}
\label{tab:recovery_length}
\renewcommand\arraystretch{1.0}
\resizebox{0.9\textwidth}{!}{%
\setlength{\tabcolsep}{0.9mm}{%
\begin{tabular}{l|ccccccccccc}
\hline
\multirow{2}{*}{Methods} & \multicolumn{5}{c}{Seq 0.8} &  & \multicolumn{5}{c}{Struct. 0.6} \\
 & Mean & Median & Short & Medium & Long &  & Mean & Median & Short & Medium & Long \\ \hline
RNAinverse & $25.92\pm1.1$ & $25.37\pm1.0$ & $25.99\pm2.0$ & $24.98\pm0.8$ & $27.54\pm1.4$ &  & $24.94\pm0.6$ & $24.24\pm0.5$ & $24.68\pm0.7$ & $24.98\pm1.0$ & $26.15\pm0.9$ \\
MCTS-RNA & $25.75\pm0.3$ & $25.61\pm0.1$ & $25.37\pm0.4$ & $26.15\pm0.5$ & $25.86\pm0.2$ &  & $25.81\pm0.5$ & $25.55\pm0.6$ & $25.38\pm0.5$ & $26.19\pm0.6$ & $25.86\pm0.9$ \\
LEARNA & $24.80\pm0.2$ & $24.55\pm0.3$ & $24.81\pm0.4$ & $24.86\pm0.2$ & $24.41\pm1.0$ &  & $24.96\pm0.2$ & $24.43\pm0.4$ & $24.88\pm0.5$ & $25.15\pm0.5$ & $24.36\pm0.6$ \\
MetaLEARNA & $29.10\pm0.6$ & $29.09\pm0.5$ & $27.43\pm1.5$ & $29.46\pm0.7$ & $32.40\pm0.9$ &  & $27.83\pm2.8$ & $27.95\pm2.5$ & $25.53\pm1.8$ & $29.51\pm0.6$ & $30.75\pm4.5$ \\
gRNAde & $42.67\pm5.3$ & $43.03\pm6.0$ & $36.25\pm2.0$ & $44.86\pm4.9$ & $46.06\pm6.1$ &  & $43.46\pm2.2$ & $43.37\pm2.7$ & $38.01\pm1.4$ & $49.82\pm2.7$ & $41.24\pm3.1$ \\
PiFold & $50.03\pm4.7$ & $50.32\pm6.0$ & $41.34\pm3.3$ & $53.20\pm3.7$ & $54.75\pm5.9$ &  & $47.89\pm5.4$ & $48.76\pm6.6$ & $40.13\pm1.0$ & $54.95\pm5.3$ & $45.62\pm7.7$ \\
StructGNN & $51.29\pm5.9$ & $52.40\pm8.0$ & $42.74\pm2.5$ & $54.45\pm7.1$ & $54.44\pm7.2$ &  & $55.20\pm6.9$ & $54.94\pm8.6$ & $46.36\pm1.0$ & $63.86\pm8.5$ & $48.48\pm11.3$ \\
GVP-GNN & $51.66\pm4.9$ & $53.48\pm6.4$ & $42.70\pm2.4$ & $56.20\pm5.7$ & $53.30\pm5.7$ &  & $53.76\pm5.4$ & $54.02\pm5.9$ & $45.80\pm0.7$ & $62.28\pm7.5$ & $47.39\pm9.0$ \\
RiboDiffusion & $\textbf{57.32}\pm4.1$ & $\textbf{58.79}\pm4.9$ & $\textbf{52.01}\pm3.1$ & $\textbf{59.95}\pm3.4$ & $\textbf{58.91}\pm5.7$ &  & $\textbf{66.50}\pm5.3$ & $\textbf{66.72}\pm5.8$ & $\textbf{61.51}\pm1.4$ & $\textbf{73.89}\pm8.4$ & $\textbf{57.98}\pm7.8$ \\ \hline
\multirow{2}{*}{Methods} & \multicolumn{5}{c}{Seq 0.6} &  & \multicolumn{5}{c}{Struct 0.5} \\
 & Mean & Median & Short & Medium & Long &  & Mean & Median & Short & Medium & Long \\ \hline
RNAinverse & $25.35\pm0.5$ & $24.30\pm0.6$ & $25.66\pm0.6$ & $25.48\pm1.8$ & $27.76\pm2.1$ &  & $25.82\pm0.6$ & $24.79\pm0.9$ & $25.38\pm1.1$ & $25.39\pm1.0$ & $27.69\pm1.6$ \\
MCTS-RNA & $25.81\pm0.2$ & $25.67\pm0.2$ & $25.29\pm0.7$ & $26.22\pm0.5$ & $26.29\pm0.6$ &  & $25.93\pm0.4$ & $25.47\pm0.4$ & $25.49\pm0.5$ & $26.28\pm0.7$ & $26.06\pm0.6$ \\
LEARNA & $24.93\pm0.1$ & $24.78\pm0.1$ & $24.92\pm0.2$ & $25.04\pm0.6$ & $24.34\pm1.0$ &  & $25.00\pm0.2$ & $24.42\pm0.6$ & $25.23\pm0.3$ & $24.64\pm0.5$ & $24.02\pm1.2$ \\
MetaLEARNA & $29.07\pm3.2$ & $29.89\pm3.0$ & $25.99\pm2.4$ & $29.81\pm0.4$ & $33.89\pm3.3$ &  & $28.13\pm3.5$ & $28.18\pm3.5$ & $25.81\pm2.0$ & $29.54\pm0.9$ & $30.87\pm3.5$ \\
gRNAde & $47.28\pm4.3$ & $49.59\pm5.4$ & $37.60\pm1.7$ & $48.66\pm8.9$ & $47.34\pm3.5$ &  & $43.36\pm6.4$ & $43.61\pm7.6$ & $36.82\pm1.5$ & $47.06\pm6.5$ & $41.74\pm9.0$ \\
PiFold & $46.74\pm2.9$ & $48.54\pm3.9$ & $37.11\pm1.6$ & $47.35\pm4.4$ & $51.32\pm5.0$ &  & $49.22\pm3.0$ & $50.06\pm3.8$ & $42.48\pm3.0$ & $53.51\pm3.6$ & $46.90\pm5.3$ \\
StructGNN & $54.23\pm4.6$ & $57.97\pm7.0$ & $41.49\pm1.7$ & $56.09\pm6.9$ & $53.32\pm11.0$ &  & $52.99\pm8.6$ & $51.81\pm10.7$ & $44.56\pm2.4$ & $59.33\pm8.3$ & $45.06\pm14.3$ \\
GVP-GNN & $54.27\pm3.9$ & $57.60\pm5.6$ & $42.54\pm1.9$ & $56.17\pm5.8$ & $54.20\pm9.2$ &  & $50.91\pm5.7$ & $50.37\pm6.9$ & $44.74\pm2.0$ & $56.51\pm7.1$ & $44.21\pm9.5$ \\
RiboDiffusion & $\textbf{59.06}\pm2.8$ & $\textbf{61.84}\pm4.2$ & $\textbf{50.68}\pm2.1$ & $\textbf{59.66}\pm4.0$ & $\textbf{59.79}\pm7.9$ &  & $\textbf{60.48}\pm6.6$ & $\textbf{59.31}\pm7.9$ & $\textbf{55.40}\pm3.8$ & $\textbf{65.69}\pm9.0$ & $\textbf{51.14}\pm10.5$ \\ \hline
\multirow{2}{*}{Methods} & \multicolumn{5}{c}{Seq 0.4} &  & \multicolumn{5}{c}{Struct 0.4} \\
 & Mean & Median & Short & Medium & Long &  & Mean & Median & Short & Medium & Long \\ \hline
RNAinverse & $25.53\pm0.7$ & $24.79\pm1.0$ & $25.29\pm0.4$ & $26.18\pm1.7$ & $27.27\pm1.8$ &  & $25.54\pm0.5$ & $24.47\pm0.6$ & $25.36\pm1.0$ & $24.94\pm0.7$ & $27.08\pm1.9$ \\
MCTS-RNA & $25.97\pm0.0$ & $25.86\pm0.2$ & $25.44\pm0.3$ & $26.48\pm0.3$ & $26.34\pm0.7$ &  & $25.81\pm0.4$ & $25.30\pm0.3$ & $25.27\pm0.5$ & $26.17\pm0.6$ & $25.86\pm1.0$ \\
LEARNA & $25.03\pm0.1$ & $24.55\pm0.3$ & $25.16\pm0.1$ & $24.84\pm0.4$ & $25.01\pm1.9$ &  & $25.05\pm0.1$ & $24.62\pm0.5$ & $25.21\pm0.2$ & $24.70\pm0.6$ & $24.02\pm1.2$ \\
MetaLEARNA & $28.94\pm1.1$ & $29.54\pm2.3$ & $25.83\pm2.2$ & $29.94\pm0.5$ & $35.36\pm2.8$ &  & $28.14\pm3.3$ & $28.31\pm3.2$ & $25.84\pm1.9$ & $29.45\pm0.6$ & $30.01\pm4.2$ \\
gRNAde & $43.58\pm7.6$ & $45.41\pm10.0$ & $36.02\pm2.6$ & $43.91\pm2.0$ & $46.84\pm12.5$ &  & $44.00\pm5.7$ & $44.10\pm7.1$ & $37.24\pm1.3$ & $48.01\pm5.6$ & $41.74\pm8.9$ \\
PiFold & $47.41\pm5.0$ & $49.00\pm6.7$ & $37.64\pm1.8$ & $50.38\pm4.7$ & $52.11\pm9.8$ &  & $49.84\pm2.7$ & $50.61\pm3.8$ & $42.39\pm2.9$ & $54.33\pm3.5$ & $45.92\pm6.3$ \\
StructGNN &  $50.40\pm6.7$ & $52.57\pm10.8$ & $41.03\pm1.5$ & $51.98\pm4.6$ & $53.33\pm13.9$ &  & $54.65\pm7.8$ & $53.98\pm9.7$ & $45.39\pm2.5$ & $61.35\pm7.0$ & $44.62\pm14.4$ \\
GVP-GNN & $50.55\pm4.7$ & $52.59\pm7.0$ & $41.77\pm0.9$ & $53.73\pm5.6$ & $51.48\pm9.3$ &  & $52.29\pm5.1$ & $51.84\pm6.6$ & $45.26\pm1.9$ & $58.26\pm5.9$ & $44.04\pm9.5$ \\
RiboDiffusion & $\textbf{57.24}\pm5.0$ & $\textbf{59.94}\pm7.7$ & $\textbf{50.06}\pm2.4$ & $\textbf{58.33}\pm4.5$ & $\textbf{58.85}\pm11.4$ &  & $\textbf{62.13}\pm6.0$ & $\textbf{61.09}\pm7.6$ & $\textbf{56.48}\pm3.9$ & $\textbf{67.94}\pm7.7$ & $\textbf{50.36}\pm10.9$ \\ \hline
\end{tabular}%
}
}
\end{table*}

We comprehensively evaluate and analyze RiboDiffusion for tertiary structure-based RNA inverse folding.
Additional results can be found in supplemental materials.
The source code is provided at
\href{https://github.com/ml4bio/RiboDiffusion}{https://github.com/ml4bio/RiboDiffusion}.

\subsection{Dataset Construction}

We gather a dataset of RNA tertiary structures from the PDB database for RNA inverse folding.
The dataset contains individual RNA structures and single-stranded RNA structures extracted from complexes.
After filtering based on sequence lengths ranging from $20$ to $280$, there is a total of $7.322$ RNA tertiary structures and $2,527$ unique sequences.
In addition to experimentally determined data, we construct augment training data by predicting structures with RhoFold \citep{shen2022e2efold}. 
The structures predicted from RNAcentral sequences \citep{rnacentral2019rnacentral} are filtered by pLDDT to keep only high-quality predictions, resulting in $17,000$ structures.

To comprehensively evaluate models, we divide the structures determined by experiments into training, validation, and test sets based on sequence similarity and structure similarity with different clustering thresholds.
We use PSI-CD-HIT \citep{fu2012cd} to cluster sequences based on nucleotide similarity. We set the threshold at $0.8/0.6/0.4$ and obtain $1,252/1,157/1,114$ clusters, respectively. 
For structure similarity clustering, we calculate the TM-score matrix using US-align \citep{zhang2022us} and apply the agglomerative clustering algorithm from scipy \citep{virtanen2020scipy} on the similarity matrix.
We achieve $2,036/1,659/1,302$ clusters with TM-score thresholds of $0.6/0.5/0.4$. 
We randomly split the clusters into three groups: $15\%$ for testing, $10\%$ for validation, and the remaining for training. We perform $4$ random splits with non-overlapping testing and validation sets for each split strategy to evaluate models.
The augmented training data is also filtered strictly based on the similarity threshold with the validation and testing sets for each split.

\subsection{RNA Inverse Folding Benchmarking}
\textit{Baselines.}
We compare our model with four machine learning baselines with tertiary structure input, including \textbf{gRNAde} \citep{joshi2023multi}, \textbf{PiFold} \citep{gao2022pifold}, \textbf{StructGNN} \citep{ingraham2019generative}, \textbf{GVP-GNN} \citep{jing2020learning}.
While gRNAde is a concurrent graph-based RNA inverse folding method, PiFold, StructGNN, and GVP-GNN are representative deep-learning methods of protein inverse folding, which are modified here to be compatible with RNA.
Implementation details of these model modifications are in the supplementary material.
These methods use the same 3-atom RNA backbone representation.
We also introduce RNA inverse folding methods with secondary structures as input for comparison.
\textbf{RNAinverse} \citep{hofacker1994fast} is an energy-based local searching algorithm for secondary structure constraints.
\textbf{MCTS-RNA} \citep{yang2017rna} searches candidates based on Monte Carlo tree search.
\textbf{LEARNA} and \textbf{MetaLEARNA} are deep reinforcement learning approaches \citep{runge2019learning} to design RNA that folds into the given secondary structures.
Each method generates a sequence for every RNA backbone for benchmarking.

\begin{figure*}[t]
\centering
\includegraphics[width=0.82\textwidth]{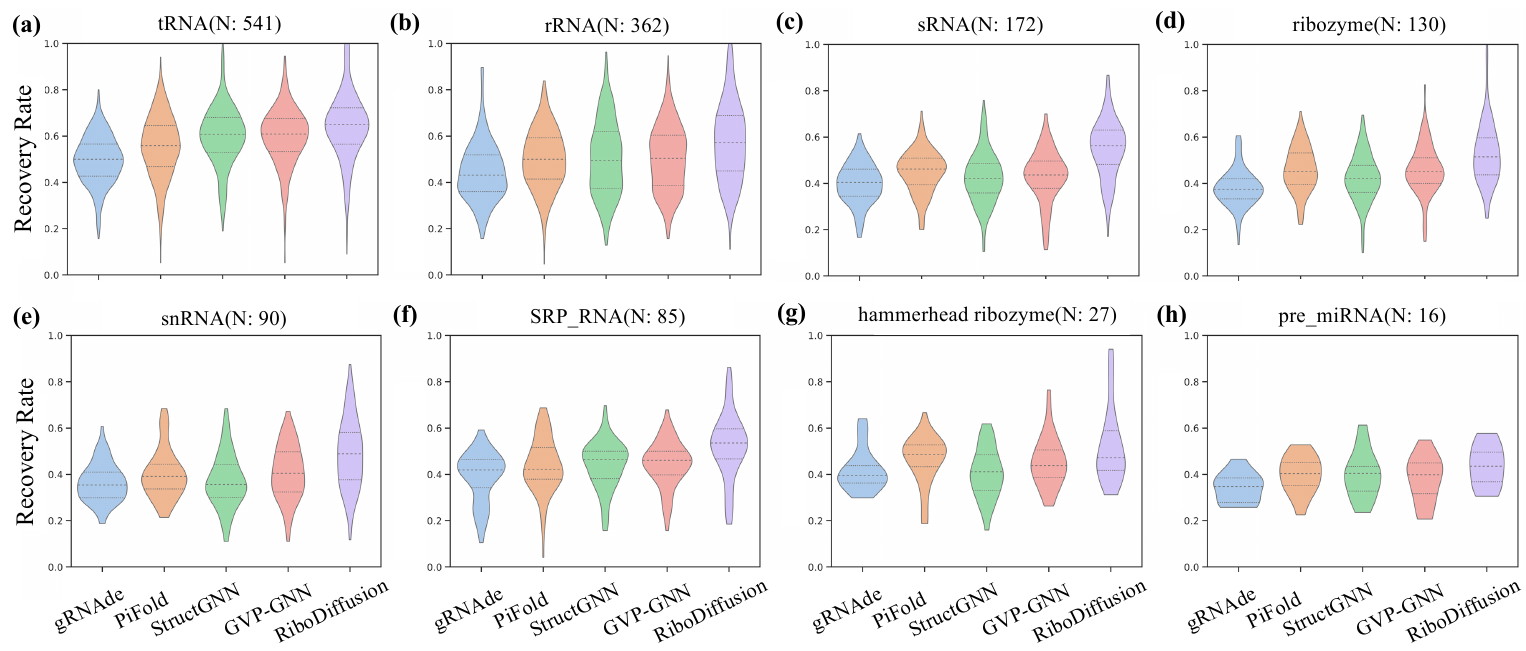}
\caption{
Violin plots for the recovery rate distribution of methods for different types of RNA, including tRNA, rRNA, sRNA, ribozyme, snRNA, SRP RNA, hammerhead ribozyme, and pre miRNA.}
\label{fig:rna_types}
\end{figure*}

\textit{Metrics.}
The recovery rate is a commonly used metric in inverse folding that shows how much of the sequence generated by the model matches the original native sequence. 
While similar sequences have a higher chance of achieving the correct fold, the recovery rate is not a direct measure of structural fitness.
We further evaluate with two metrics: the F1 Score, which assesses the alignment between the generated sequence's predicted secondary structure (via RNAfold \citep{gruber2008vienna}) and the secondary structure extracted from the input's tertiary structure (using DSSR \citep{lu2015dssr}), and the success rate determined by Rfam's covariance model \citep{kalvari2021rfam}, which evaluates the preservation of family-specific information in the generated sequences, indicating conserved structures and functions. Average success rates across families are reported.

\begin{table}[t]
\centering
\caption{Comparison of secondary structure similarity and success rate of family preservation. F1: F1 score. Suc.: success rate of family preservation.}
\label{tab:bench2}
\renewcommand\arraystretch{1.0}
\resizebox{0.95\columnwidth}{!}{%
\setlength{\tabcolsep}{0.5mm}{%
\begin{tabular}{llccccc}
\hline
 &  & gRNAde & PiFold & StructGNN & GVP-GNN & RiboDiffusion \\ \hline
\multirow{2}{*}{Seq 0.8} & F1 & 0.564 & 0.408 & 0.761 & 0.765 & 0.744 \\
 & Suc. & 0.035 & 0.100 & 0.266 & 0.268 & 0.370 \\ \hline
\multirow{2}{*}{Seq 0.6} & F1 & 0.142 & 0.336 & 0.709 & 0.740 & 0.749 \\
 & Suc. & 0.018 & 0.031 & 0.217 & 0.186 & 0.316 \\ \hline
\multirow{2}{*}{Seq 0.4} & F1 & 0.424 & 0.388 & 0.777 & 0.802 & 0.785 \\
 & Suc. & 0.033 & 0.033 & 0.164 & 0.138 & 0.224 \\ \hline
\multirow{2}{*}{Str 0.6} & F1 & 0.571 & 0.434 & 0.774 & 0.785 & 0.856 \\
 & Suc. & 0.036 & 0.023 & 0.206 & 0.163 & 0.305 \\ \hline
\multirow{2}{*}{Str 0.5} & F1 & 0.731 & 0.440 & 0.763 & 0.766 & 0.786 \\
 & Suc. & 0.064 & 0.028 & 0.140 & 0.150 & 0.195 \\ \hline
\multirow{2}{*}{Str 0.4} & F1 & 0.738 & 0.428 & 0.744 & 0.761 & 0.790 \\
 & Suc. & 0.060 & 0.031 & 0.128 & 0.134 & 0.l77 \\ \hline
\end{tabular}%
}
}
\end{table}

We present recovery rate results in \autoref{tab:recovery_length}, which contains the average and standard deviation of four non-overlapping test sets for each model in different cluster settings.
Our model outperforms the second best method by $11\%$ on average for sequence similarity splits and $16\%$ for structure similarity splits.
RiboDiffusion consistently achieves better recovery rates in RNA with varying degrees of sequence or structural differences from training data.
Methods based on tertiary structures outperform those based on secondary structures, as the latter contains less structural information.
Extra results are shown in \autoref{tab:bench2}. 
It is worth noting that the tools used in these two metrics may contain errors. Our proposed method outperforms or matches the baseline methods in secondary structure alignments and more effectively retains family information from the input RNA.


\begin{figure}[t]
\centering
\includegraphics[width=0.85\columnwidth]{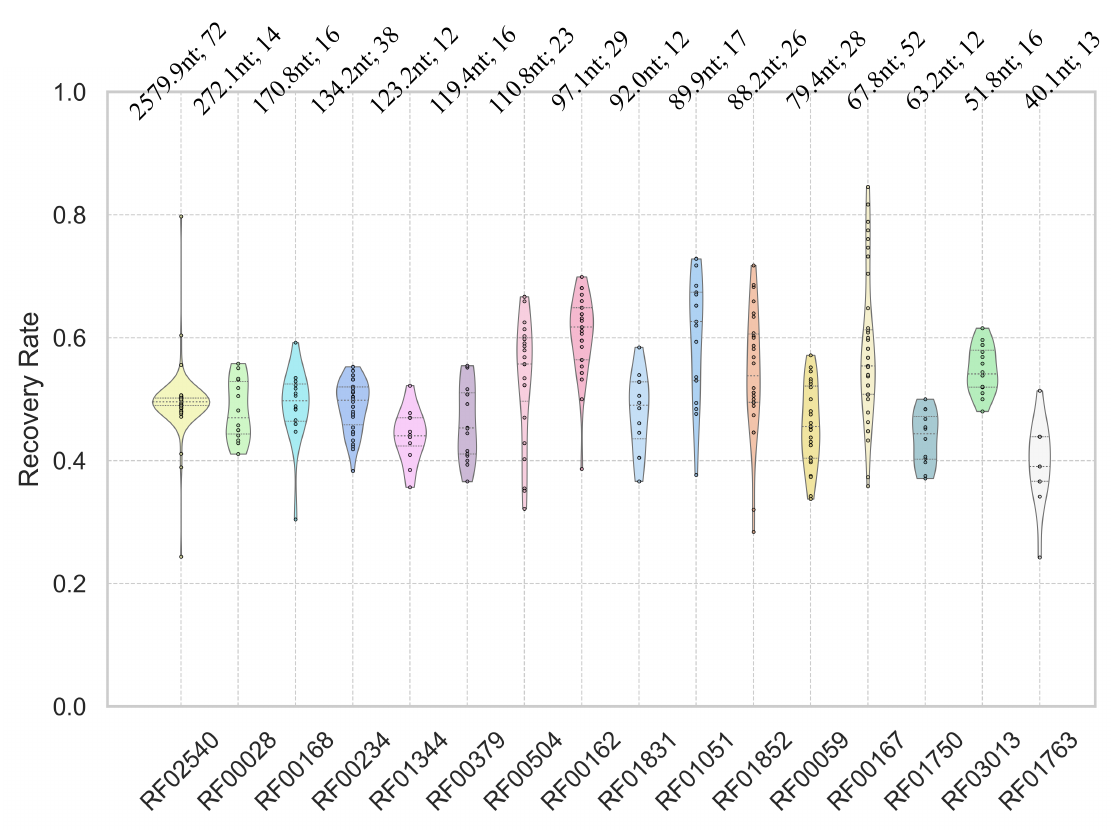}
\caption{Performance of RiboDiffusion on different RNA families under the cross-family setting. The average length and number of tertiary structures for each family are marked above violin plots.
}
\label{fig:cross_rfam}
\end{figure}

We further classify the RNA in the test set based on its length and type to compare the model performance differences more thoroughly.
First, we divide RNA into three categories based on the number of nucleotides (nt), \ie Short ($50$ nt or less), Medium (more than $50$ nt but less than $100$ nt), and Long ($100$ nt or more).
It can be observed in \autoref{tab:recovery_length} that RiboDiffusion maintains performance advantages across different lengths of RNA.
Short RNAs present a challenge for the model to recover the original sequence due to their flexible conformation, causing a relatively low recovery rate when compared to medium-length RNAs.
A more detailed correlation of RiboDiffusion performance with RNA length is shown in supplemental materials.
Each split shows similar patterns: RiboDiffusion has higher variance in short RNA inverse folding, and the model's performance becomes limited as RNA length increases.
Moreover, \autoref{fig:rna_types} shows the recovery rate distribution of different RNA types with over $10$ structures in test sets, including rRNA, tRNA, sRNA, ribozymes, etc.\ 
The RNA type information is collected from \citep{rnacentral2019rnacentral}.
Compared to other baselines, RiboDiffusion still has a better recovery rate distribution across RNA types.
Through comprehensive benchmarking, we have observed remarkable performance improvement in tertiary structure-based RNA inverse folding achieved by RiboDiffusion.

\subsection{Analysis of RiboDiffusion}

We dive into a more comprehensive analysis of RiboDiffusion.

\begin{figure*}[!htbp]

\centering

\includegraphics[width=0.89\textwidth]{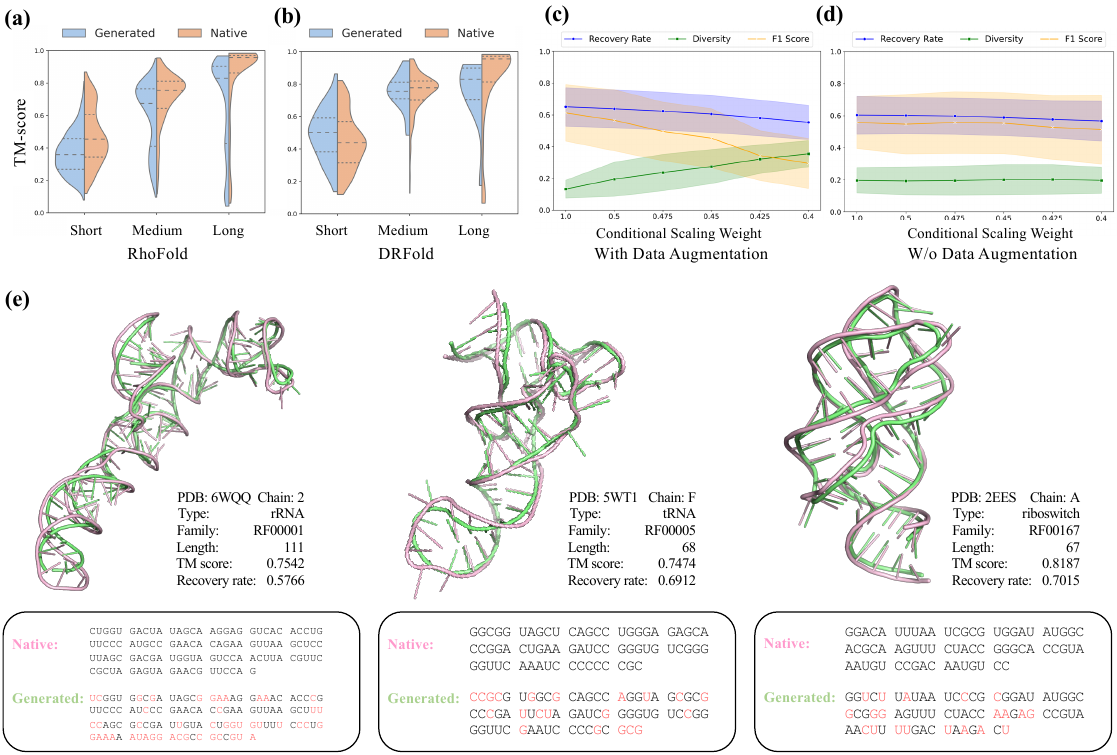}

\caption{
\textbf{Analysis of RiboDiffusion.}
\textbf{(a)-(b)} In-silico folding validation results that show the TM-score between structures predicted by RhoFold or DRFold and the given fixed RNA backbones (on \textit{Seq. 0.4} split).
\textit{Native} represents structures predicted from original sequences of given backbones as references, while \textit{Generated} represents structures predicted from generated sequences. 
\textbf{(c)-(d)} Trade-offs between the diversity of generated sequences and recovery rate,  as well as refolding F1-score (including models with and without augmented data).
\textbf{(e)} Visualization of input RNA structures (pink) and predicted structures (green) of generated sequences. 
The generated sequences and the corresponding native sequences are shown below the structure visualization, where different nucleotide types are marked in red.
}

\label{fig:fold_exp}
\end{figure*}

\textit{Cross-family performance.}
We repartition the dataset with the cross-family setting to further verify the generalization of our model. 
We obtain the RNA family corresponding to the tertiary structure from \citep{kalvari2021rfam}, then randomly select four families for testing and others for training.
The experimental results of $4$ non-overlapping splits are shown in \autoref{fig:cross_rfam}.
The average recovery rate of RiboDiffusion in each family generally ranges between $0.4$ and $0.6$. Especially, our model performs well on RF02540 whose sequence length far exceeds the training set.
Although the performance is slightly worse than other splits in \autoref{tab:recovery_length}, these results still illustrate that our model can handle RNA families that do not appear in the training data, considering that cross-family is inherently a more difficult setting.

\textit{In-silico tertiary structure folding validation.} 
To verify whether RiboDiffusion generated sequences can fold into a given RNA 3D backbone, we use computational methods to predict RNA structures (\ie RhoFold \citep{shen2022e2efold} and DRFold \citep{li2023integrating}) to obtain their tertiary structures.
Structure prediction models with a single sequence input are used due to the difficulty in finding homologous sequences for generated sequences and performing multiple sequence alignment.
We take the TM-score of C1' backbone atoms to measure the similarity between the predicted RNA structure of generated sequences and the given fixed backbones.
Note that in-silico folding validation contains two sources of errors. One is the structure prediction error of the folding method itself, and the other is the sequence quality generated by RiboDiffusion.
Therefore, we also predict the structure from the original native sequence using the same folding method and compare it to the given RNA backbone as an error and uncertainty reference. 

As depicted in \autoref{fig:fold_exp} (a), sequences generated by RiboDiffusion exhibit promising folding results in the fixed backbone for medium-length and long-length RNAs.
However, the performance for short-length RNAs is relatively poor, which is affected by the unsatisfied recovery rate of our model and the limitations of RhoFold itself.
We also show the folding performance using DRFold in \autoref{fig:fold_exp} (b), where RiboDiffusion exhibits distribution shapes similar to those of using RhoFold.
Here, due to the limitation of DRFold inference speed, we only test on the representative sequence of each cluster instead of the entire test set.
We further make in-silico folding (with RhoFold) case studies of rRNA, tRNA, and riboswitch in \autoref{fig:fold_exp} (e).
RiboDiffusion generates new sequences that are different but still tend to fold into similar geometries.
To alleviate concerns about the independence of structure prediction and inverse folding models, we provide results from alternative tools and evaluations of structures independent of current datasets as an extra reference in the supplementary material.

\textit{Trade-off between sequence recovery and diversity.}
Exploring novel RNA sequences that have the potential to collapse into a fixed backbone distinct from native sequences is a realistic demand for RNA design. However, there is a trade-off between the diversity and recovery rate of the generated sequences. 
RiboDiffusion can achieve this balance by controlling the conditional scaling weight.
For the representative input backbone of each cluster, we generate $8$ sequences in total to report diversity.
The diversity within the generated set of sequences $G$ is defined as $\mathrm{IntDiv}(G) = 1 - \frac{1}{|G|^2}\sum_{S_1, S_2 \in G}\mathrm{Sim}(S_1, S_2)$ \citep{benhenda2017chemgan}. The function $\mathrm{Sim}$ compares two sequences by calculating the ratio of the length of the aligned subsequence to the length of the shorter sequence.
In \autoref{fig:fold_exp} (c), it is evident that the mean diversity of generated sequences in the test sets begins to increase when the conditional scaling weight is set to $0.5$, while the recovery rate and the F1 score decrease to some extent.
Therefore, we recommend using a value between $0.5$ and $0.35$ to adjust the sequence diversity.


\textit{Training data augmentation analysis.} 
Augmenting training data is primarily driven by the scarcity and limited diversity of RNA available in PDB.
\autoref{tab:aug} indicates that the incorporation of additional RhoFold predictions improves the overall generated sequence quality. This augmentation also enhances the adjustment ability of RiboDiffusion for sequence diversity, as shown in \autoref{fig:fold_exp} (d), where the sequence diversity of the model without the augmented data remains relatively low.
Notably, the noisy nature of augmented data requires appropriate preprocessing and filtering for quality assurance.

\begin{table}[!htbp]
\centering
\caption{Ablation study on data augmentation. Rec.: recovery rate.}
\label{tab:aug}
\resizebox{0.95\columnwidth}{!}{%
\setlength{\tabcolsep}{0.6mm}{%
\begin{tabular}{lcccc}
\hline
 & Rec. Mean & Rec. Median & F1 score & Rfam Success \\
RiboDiffusion & 57.24\% & 59.94\% & 0.785 & 0.224 \\
w/o Augment & 55.26\% & 57.01\% & 0.768 & 0.221 \\ \hline
\end{tabular}%
}
}
\end{table}


\section{Conclusion}

We propose RiboDiffusion, a generative diffusion model for RNA inverse folding based on tertiary structures.
By benchmarking methods on sequence and structure similarity splits, comparing performance across RNA length and type, and validating with in-silico folding, we demonstrate the effectiveness of our model.
Our model can also make trade-offs between recovery and diversity, and handle cross-family inverse folding.
In future work,
we aim to expand the scope of RiboDiffusion by exploring RNA sequences that span larger magnitudes in size and integrate contact information from the complex into the model.
Our ultimate objective is to utilize the model for designing functional RNA like ribozymes, riboswitches, and aptamers, and to verify its effectiveness in wet lab experiments.


\bibliographystyle{abbrvnat}
\bibliography{Li.10}

\end{document}


\journaltitle{Bioinformatics}
\DOI{https://doi.org/xxx}
\copyrightyear{xxxx}
\appnotes{\bf ISMB 2024}

\firstpage{1}

\title{RiboDiffusion: Tertiary Structure-based RNA Inverse Folding with Generative Diffusion Models}

\author[1,2,$\dagger$]{Han Huang}
\author[1,3,$\dagger$]{Ziqian Lin}
\author[1]{Dongchen He}
\author[1]{Liang Hong}
\author[1,$\ast$]{Yu Li}

\authormark{Huang, H., et al.}

\address[1]{Department of Computer Science and Engineering, CUHK, Hong Kong SAR, China}
\address[2]{School of Computer Science and Engineering, Beihang University, Beijing, China}
\address[3]{Nanjing University, Nanjing, China}


\abstract{}

\maketitle

\begin{appendices}

\section{Dataset}

\begin{figure*}[!htbp]
\centering
\includegraphics[width=0.8\textwidth]{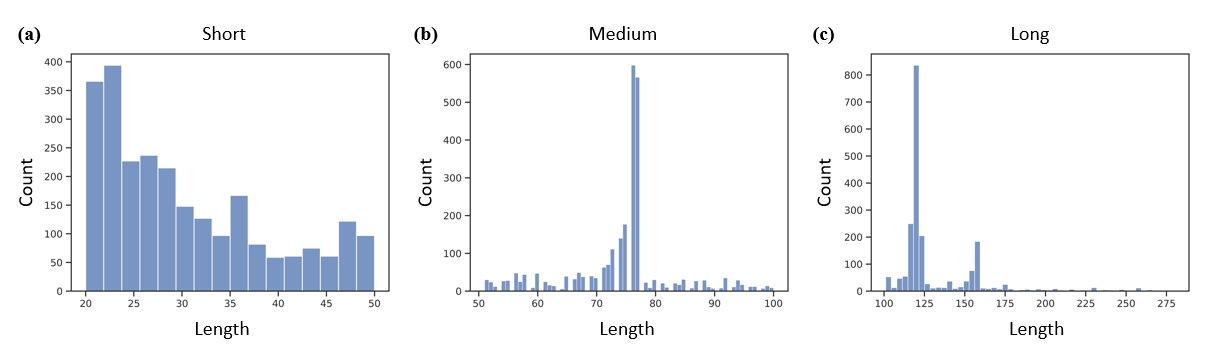}
\caption{\textbf{Length distribution of the experimentally determined structures.} \textbf{(a)-(c)} Length distribution of short ($L<=50\text{nt}$), medium ($50\text{nt}<L<=100\text{nt}$) and long ($L>100\text{nt}$) RNA.}
\label{fig:dataset_length}
\end{figure*}

Our dataset has $7,322$ experimentally determined RNA 3D structures. 
We provide the length histogram of these structures in Figure \ref{fig:dataset_length}.

\section{Experiment Details and Results}

\subsection{Secondary Structure-based Methods}

We apply several RNA secondary structure-based inverse folding and protein inverse folding methods to compare model performance. 
For RNA secondary structure inverse folding methods, we extract secondary structures in dot-bracket form through DSSR \citep{lu2015dssr}. We use the default optimized hyperparameters of these methods. For LEARNA and MetaLEARNA \citep{runge2019learning}, we set the design time limit to $600$ seconds.

\subsection{Protein Inverse Folding Methods}

GVP-GNN \citep{jing2020learning}, PiFold \citep{gao2022pifold} and StructGNN \citep{ingraham2019generative} are models based on graph neural networks which are first used for protein inverse folding methods. These methods can well extract geometric features using their graph neural network module. As a result, we construct RNA geometric features as model input.

For GVP-GNN, we use the same input features of RiboDiffusion as RiboDiffusion has a structure module based on GVP-GNN. The scalar node features contain dihedral angles of each nucleotide. The vector node features consist of forward and reverse vectors of sequential C1’ atoms, as well as the local orientation vectors of C1’ to C4’ and N1/N9.  The initial embedding of each edge consists of its connected C1’ atom’s direction vector, Gaussian radial basis encoding for their Euclidean distance, and sinusoidal position encoding of the relative distance in the sequence. 

StructGNN consists of two parallel encoders to obtain embeddings of substructures and molecules, followed by a feed-forward neural network for prediction. 
PiFold contains PiGNN layers considering multi-scale residue interactions in node, edge, and global context levels of the graph and a linear layer. 
For PiFold and StructGNN, we construct distance, angle, and direction features for single or paired nucleotides similar to those in protein. The scalar node features contain dihedral angles of each nucleotide and Gaussian radial basis encoding for every atom pair among C4', C1', N1/N9 of each nucleotide. The vector node features consist of the local orientation vectors of C1’ to C4’ and N1/N9. The scalar edge features contain Gaussian radial basis encoding of every atom pair among C4', C1', N1/N9 of two different nucleotides, as well as quaternions of relative rotation between their local coordinate systems. The vector edge features consist of the orientation vectors of C1’ of one nucleotide to C4’ and N1/N9 in a different nucleotide. In these features, C4', C1', N1/N9 in nucleotide correspond to N, $\text{C}_\alpha$, C in protein residues.

Protein inverse folding methods exploit the geometric features of protein molecules. By constructing similar geometric features in our 3-atom RNA backbones, we can retrain these models on the RNA dataset. As a result, these methods can be applied to the RNA inverse folding problem.

\begin{figure}[!htbp]
\centering
\includegraphics[width=0.7\columnwidth]{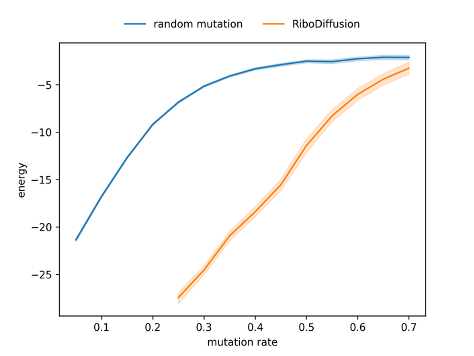}
\caption{\textbf{The correlation between different mutation rates and free energy} (with random mutation and RiboDiffusion).}
\label{fig:appendix_mut}
\end{figure}

\subsection{Metric}

\textit{Recovery rate.} 
This metric evaluates the quality of inverse folding from the perspective of sequence similarity. 
It is not perfect because it cannot directly characterize the possibility of sequences folding into a specified structure, but it still has a certain reference value.
We plot the random mutation ratio versus the free energy of the sequence folding into a given secondary structure (extracted from the tertiary structure) in Figure \ref{fig:appendix_mut}. Folding into the structure is more likely when the recovery rate is relatively high. Moreover, our method has lower free energy than random mutation at the same recovery rate.

\textit{F1 Score for secondary structure alignment.}
F1 Score is defined between the secondary structure of the generated sequence predicted by RNAfold \citep{gruber2008vienna} and the secondary structure extracted from the input tertiary structure. 
This metric reflects whether the generated sequence satisfies the folding constraints from the secondary structure level. 
However, since the secondary structures derived from both methods may have errors, we remove data that may have large errors based on the F1 score of the native sequence with a threshold of $0.7$.

\textit{Rfam success rate.}
We use Rfam's covariance model to evaluate whether the sequence obtained by inverse folding maintains the same family information with the original RNA. 
Sequences within the curated family are generally considered to have conserved structures and similar functions. 
It is also of significance to discover such new sequences through inverse folding. 
Specifically, we define the success case as whether the bit score of the generated sequence is larger than the gathering threshold.

\subsection{Hyperparameters}
Here we list the main hyperparameters we used in our model.
We construct nucleotide graphs with top-$10$ neighbors and stack $4$ layers for the graph neural network in the structure module, where the node feature dimension is $512$ and the edge feature dimension is $128$.
For the sequence module consisting of $8$ blocks, we keep the $512$ dimensions and use $8$ attention heads.
Our model is trained $40$ epochs with the learning rate $0.0002$.

\subsection{Extra In-silico Tertiary Structure Folding Results}

\begin{figure}[!htbp]
\centering
\includegraphics[width=\columnwidth]{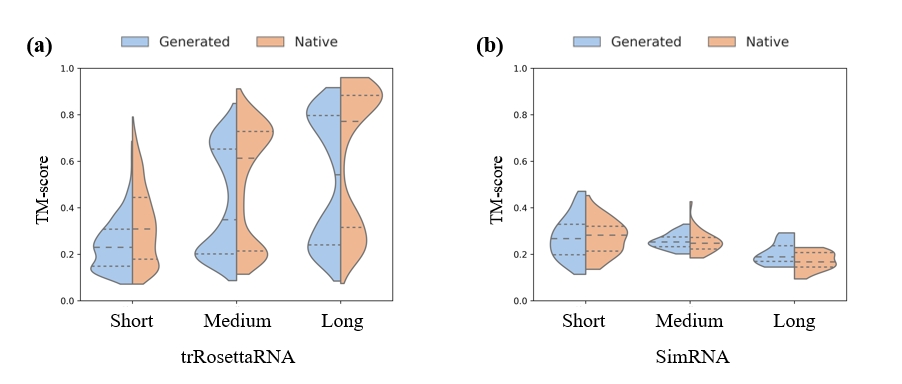}
\caption{\textbf{In-silico folding validation results of trRosettaRNA and simRNA.} In-silico folding validation results that show the TM-score between structures predicted by trRosettaRNA or simRNA and the given fixed RNA backbones (on \textit{Seq. 0.4} split).
\textit{Native} represents structures predicted from original sequences of given backbones as references, while \textit{Generated} represents structures predicted from generated sequences.}
\label{fig:appendix_fold}
\end{figure}

To alleviate concerns about the independence of structure prediction tool and inverse folding models, we use two extra computational tools, trRosettaRNA \citep{wang2023trrosettarna} and SimRNA \citep{boniecki2016simrna}, to obtain tertiary structures of generated RNA sequences. We also use these tools to predict tertiary structures from the original native sequences. 
As depicted in Figure \ref{fig:appendix_fold}(a), generated and native sequences have similar TM-score distribution when predicted by trRosettaRNA. The result of SimRNA is shown in Figure \ref{fig:appendix_fold}(b). The performance of SimRNA is relatively poor, which indicates that although generated sequences have a similar TM-score distribution to natural sequences, the refolding evaluation based on SimRNA may have a large error and uncertainty.

\begin{figure*}[!htbp]
\centering
\includegraphics[width=0.9\textwidth]{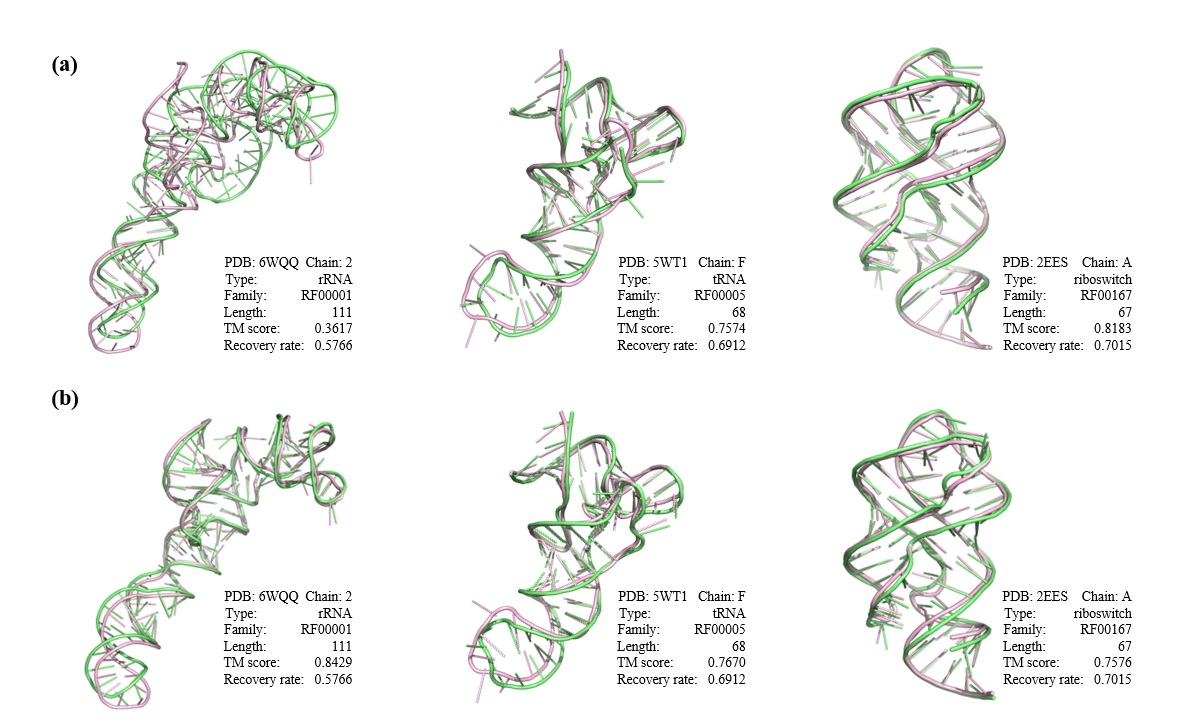}
\caption{\textbf{In-silico folding visualized results of DRFold and trRosettaRNA .}\textbf{(a)} Visualization of input RNA structures (pink) and predicted structures (green) of RiboDiffusion-DRFold pipeline. \textbf{(b)} Visualization of input RNA structures (pink) and predicted structures (green) of RiboDiffusion-trRosettaRNA pipeline.
}
\label{fig:appendix_dr_tr}
\end{figure*}

Besides RhoFold \citep{shen2022e2efold}, we also provide 3D visualized results of DRFold \citep{li2023integrating} and trRosettaRNA, which are shown in Figure \ref{fig:appendix_dr_tr}.

\subsection{Results on New RNA Structures}

We evaluate the newly published RNA structures between 2023 and 2024 as an additional reference for our model. 
After removing redundancy and removing RNAs similar to the training set, we present 8 structures that have not been trained by RiboDiffusion and RhoFold. 
The result is displayed in Table \ref{tab:pdb2023}.

\subsection{Performance on CASP15}

\begin{figure}[!htbp]
\centering
\includegraphics[width=\columnwidth]{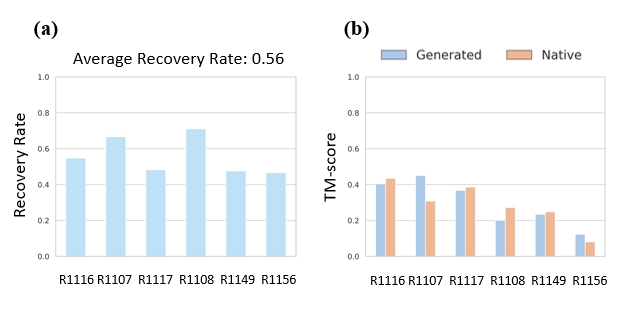}
\caption{\textbf{Performance on CASP15.}\textbf{(a)} A bar chart shows the recovery rate of RiboDiffusion on six natural RNAs in CASP15. 
\textbf{(b)}  
A bar chart displays the TM-score between predicted structures of RiboDiffusion-generated sequences and given RNA backbones. 
The TM-score of predicted structures from native sequences is displayed as a reference.}
\label{fig:appendix_CASP}
\end{figure}

To assess the generalizability of the model, RiboDiffusion is tested on six natural RNAs in CASP15 without any overlap with the training set.
As shown in Figure \ref{fig:appendix_CASP} (a) and (b), the performance of RiboDiffusion in complex RNA backbone structures is impressive, which is demonstrated by an average recovery rate of 0.56. Furthermore, the TM-score values of generated sequences are similar to the native sequences. However, it is important to note that the results of in-silico folding on CASP15 need more follow-up validation, as the TM-score value used as a reference is not satisfactory.

\subsection{Ablation Studies}
We perform additional ablation studies to validate the necessity of the sequence module.
We train the models in a sequence similarity split and a structure similarity split and report the results in Table \ref{tab:my_abl}. In our diffusion model formulation, adding the sequence module facilitates performance improvement.

\begin{table}[!htbp]
    \centering
    \caption{Mean recovery rate (\%) on ablation studies.}
    \renewcommand\arraystretch{1.2}
    \setlength{\tabcolsep}{1mm}{%
    \begin{tabular}{l|ccc}
        \hline
        Method &  \textit{Seq.} & \textit{Struct.} \\ \hline
        RiboDiffusion & 58.96 & 66.40 \\ \hline
        RiboDiffusion w/o seq  & 57.82 & 64.26 \\ \hline
    \end{tabular}
    }
    
    \label{tab:my_abl}
\end{table}

\subsection{Running Time and Scalability Analysis.}

\begin{figure}[!htbp]
\centering
\includegraphics[width=0.5\columnwidth]{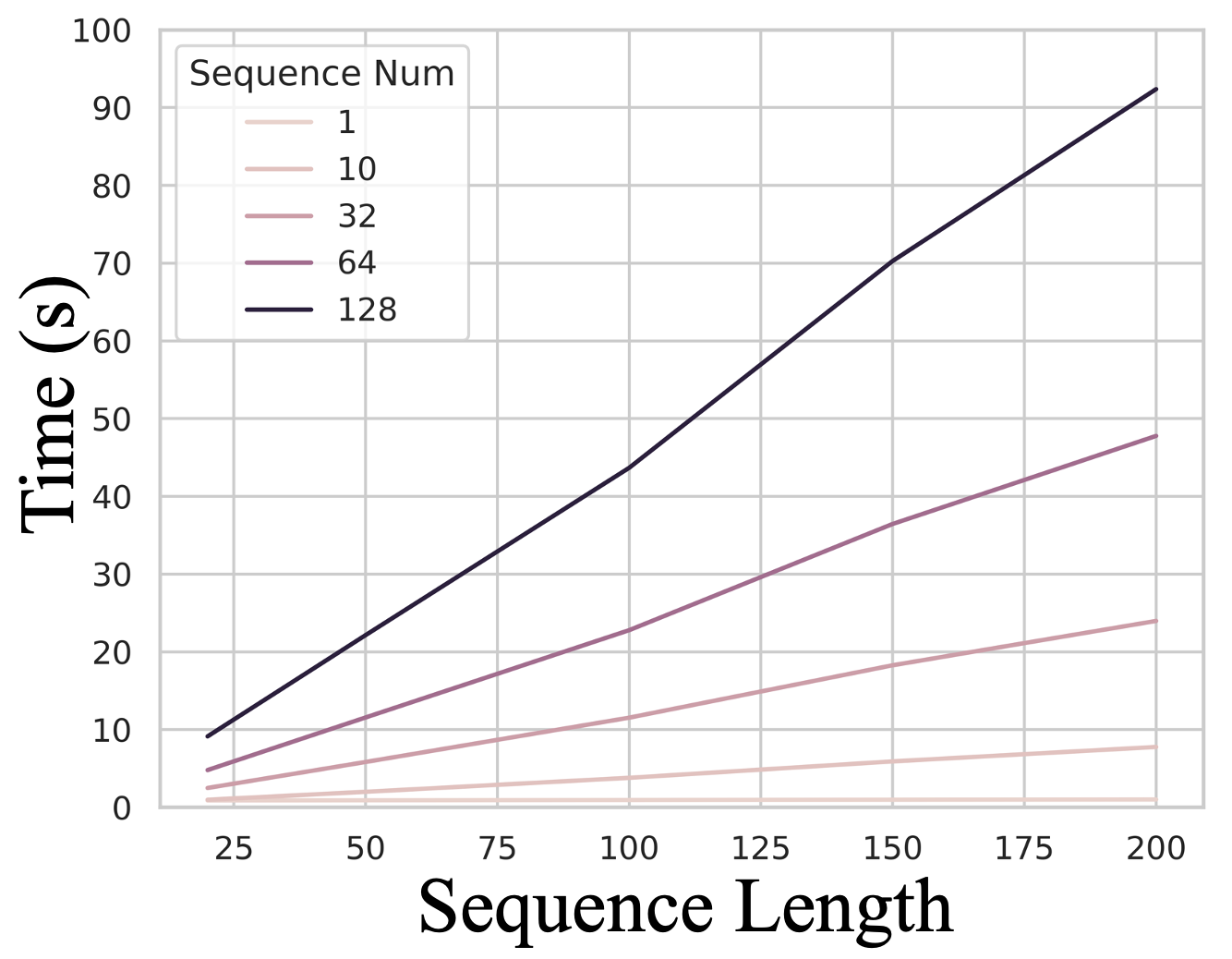}
\caption{\textbf{Running time and scalability analysis. }A line chart shows the relationship between running time and RNA sequence length when predicting different numbers of RNA sequences simultaneously.
}
\label{fig:appendix_time}
\end{figure}

The inference time of diffusion-based models is largely dependent on the number of steps in the sampling process. For the run-time analysis, we use $50$ steps identical to those in our other experiments.
On a GeForce RTX 3090 GPU, we report wall clock times of RiboDiffusion generation with different lengths of RNA and different numbers of sequences generated simultaneously in Figure \ref{fig:appendix_time}.
RiboDiffusion can finish the inverse folding of $200$ nt RNA in just one second when generating a sequence. However, when generating $128$ sequences simultaneously, RiboDiffusion experiences a significant increase in processing time, leading to limitations in scalability. We believe that the running speed of RiboDiffusion can be further improved in the future by accelerating the diffusion models, which is currently an emerging topic in machine learning.

\subsection{Extra Results on Secondary Structure Based Methods}
We report extra results of secondary structure-based inverse folding methods in Table \ref{tab:bench2_ss}.
These methods obtain high F1 scores because they directly use energy optimization to obtain sequences, making it unfair to compare with other methods. 
It is difficult for secondary structure-based inverse folding methods to generate new sequences in the same family due to the information loss compared to the tertiary structure input, even for tRNA with a more conservative shape.

\begin{table}[!htbp]
\centering
\caption{Comparison of secondary structure similarity and success rate of family preservation. The F1 score is an unfair metric for energy-optimized methods.}
\label{tab:bench2_ss}
\renewcommand\arraystretch{1.0}
\resizebox{0.8\columnwidth}{!}{%
\setlength{\tabcolsep}{0.7mm}{%
\begin{tabular}{llcccc}
\hline
 &  & rnainverse& MCTS& learna& metalearna\\ \hline
\multirow{2}{*}{Seq 0.8} & F1*& 0.990& 0.918& 0.750& 0.905\\
 & Suc. & 0.000& 0.000& 0.000& 0.000\\ \hline
\multirow{2}{*}{Seq 0.6} & F1* & 0.991& 0.922& 0.764& 0.916\\
 & Suc. & 0.000& 0.000& 0.000& 0.000\\ \hline
\multirow{2}{*}{Seq 0.4} & F1* & 0.987& 0.916& 0.796& 0.928\\
 & Suc. & 0.000& 0.000& 0.000& 0.000\\ \hline
\multirow{2}{*}{Str 0.6} & F1* & 0.990& 0.915& 0.776& 0.913\\
 & Suc. & 0.000& 0.000& 0.000& 0.000\\ \hline
\multirow{2}{*}{Str 0.5} & F1* & 0.985& 0.900& 0.789& 0.919\\
 & Suc. & 0.000& 0.000& 0.000& 0.000\\ \hline
\multirow{2}{*}{Str 0.4} & F1* & 0.987& 0.901& 0.762& 0.911\\
 & Suc. & 0.000& 0.000& 0.000& 0.000\\ \hline
\end{tabular}%
}
}
\end{table}

\subsection{Results on Remaining Dataset Splits}
\begin{figure*}[!htbp]
\centering
\includegraphics[width=\textwidth]{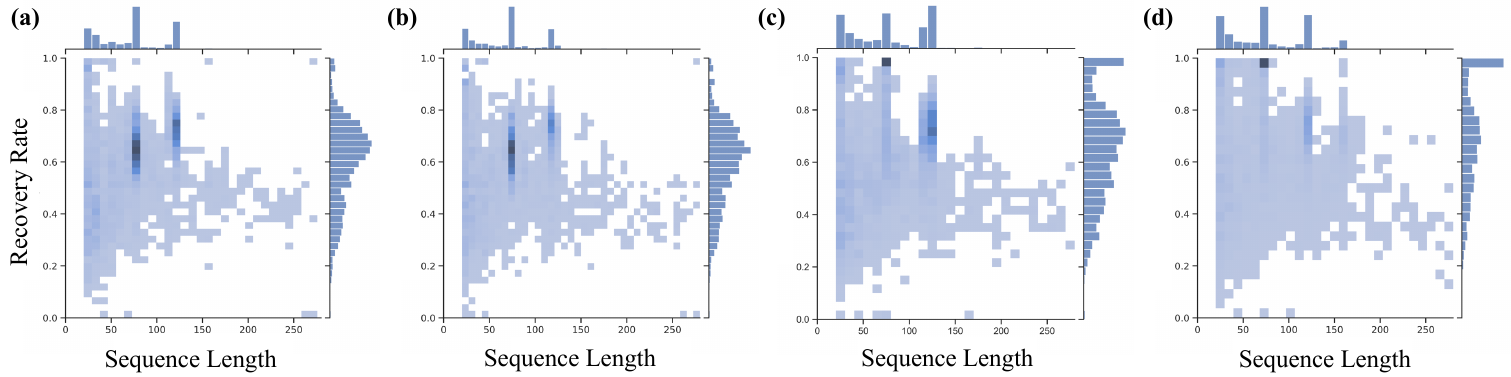}
\caption{\textbf{Bivariate distribution of sequence length and recovery rate for RiboDiffusion.} \textbf{(a)-(d)} Four joint histplots of the bivariate distribution between sequence length and recovery rate on test set splits including \textit{Seq. 0.6}, \textit{Seq. 0.8}, \textit{Struct. 0.5} and \textit{Struct. 0.6}. }
\label{fig:appendix_joint}
\end{figure*}

\begin{figure*}[!htbp]
\centering
\includegraphics[width=0.65\textwidth]{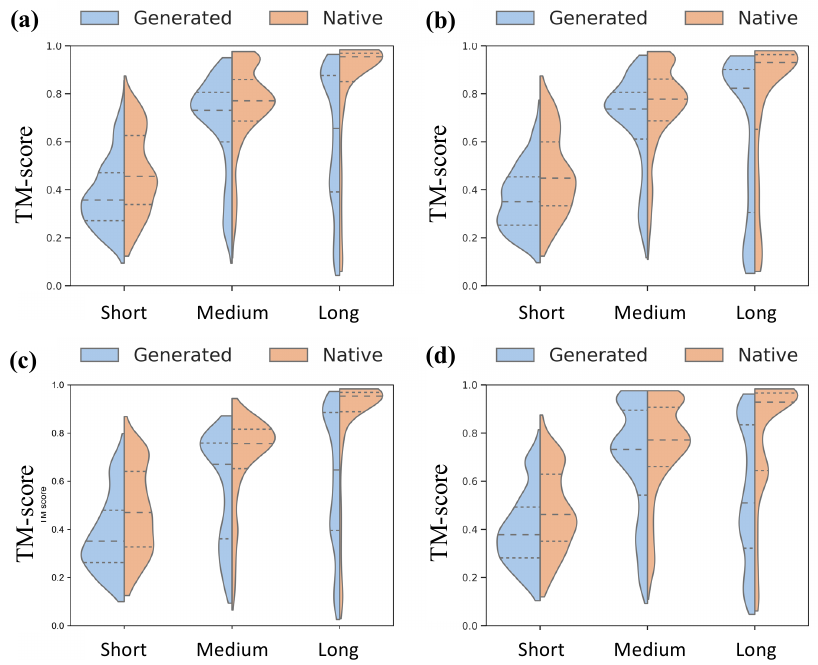}
\caption{\textbf{TM-score performance of RiboDiffusion-RhoFold pipeline about RNA length.} \textbf{(a)-(d)} Four violin plots compare the TM-score of structures predicted by RhoFold between generated RNA and native RNA on short, medium, and long RNA data in test set splits including \textit{Seq. 0.6}, \textit{Seq. 0.8}, \textit{Struct. 0.5} and \textit{Struct. 0.6}.}
\label{fig:appendix_length}
\end{figure*}

\begin{figure*}[!htbp]
\centering
\includegraphics[width=0.65\textwidth]{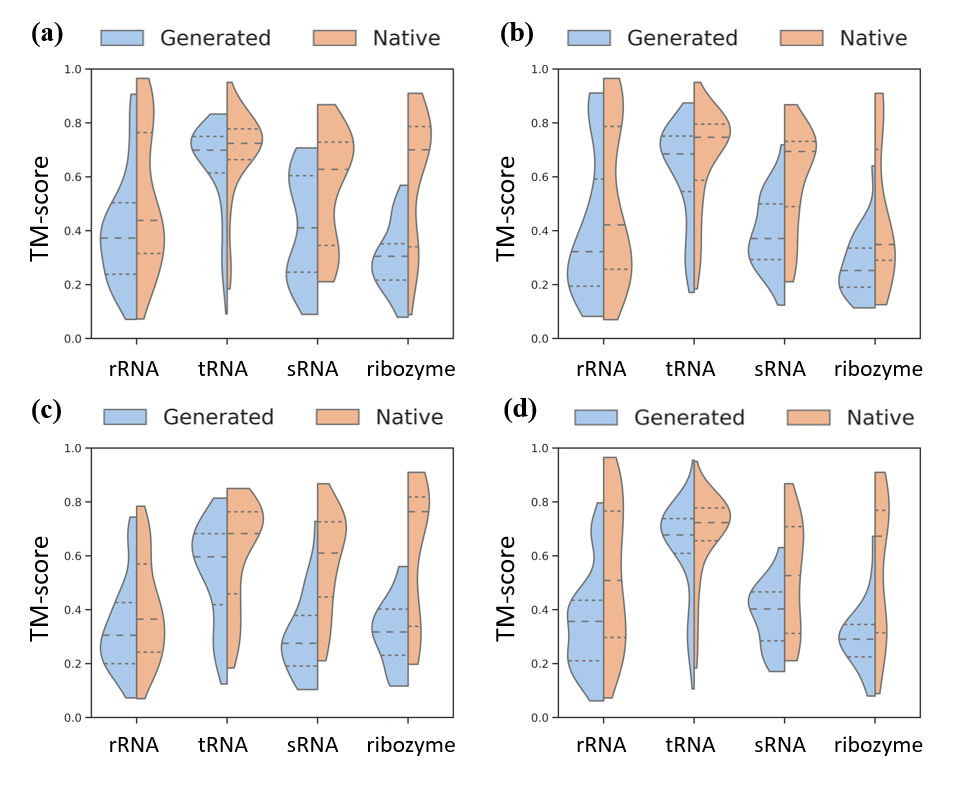}
\caption{\textbf{TM-score performance of RiboDiffusion-RhoFold pipeline about RNA type.} \textbf{(a)-(d)} Four violin plots compare the TM-score of structures predicted by RhoFold between generated RNA and native RNA on different types of RNA including rRNA, tRNA, sRNA, and ribozyme in test set splits including \textit{Seq. 0.6}, \textit{Seq. 0.8}, \textit{Struct. 0.5} and \textit{Struct. 0.6}.}
\label{fig:appendix_types}
\end{figure*}

Extra results of different dataset splits are shown in Figure 
\ref{fig:appendix_joint},  \ref{fig:appendix_length}, \ref{fig:appendix_types}. 
We show the bivariate distribution of sequence length and recovery rate for RiboDiffusion on test set splits including \textit{Seq. 0.6}, \textit{Seq. 0.8}, \textit{Struct. 0.5} and \textit{Struct. 0.6} in Figure \ref{fig:appendix_joint}. We provide additional violin plots displaying the TM-score performance of RiboDiffusion-RhoFold pipeline about RNA length on test set splits including \textit{Seq. 0.6}, \textit{Seq. 0.8}, \textit{Struct. 0.5} and \textit{Struct. 0.6} in Figure \ref{fig:appendix_length}. In Figure \ref{fig:appendix_types}, four different types of RNA are tested to evaluate the performance of RiboDiffusion. The results show that RiboDiffusion performs better on tRNA compared to rRNA. However, its performance in sRNA and ribozyme may be limited due to the scale of the relevant training data.

\begin{table*}[!htbp]

\centering
\caption{
\textbf{Results on newly published RNA structures.} TM-score (generated) is calculated between the given structure and the refolded structure from the RiboDiffusion-RhoFold pipeline. TM-score (native) is calculated between the given structure and the predicted structure of RhoFold with the original native sequence. 
}
\label{tab:pdb2023}
\renewcommand\arraystretch{1.0}
\resizebox{0.95\textwidth}{!}{%
\begin{tabular}{|c|c|c|c|l|}
\hline
\textbf{PDB id} & \textbf{Recovery rate} & \textbf{TM-score (generated)} & \textbf{TM-score (native)} & \textbf{Generated sequence}                                                                                                                                                                                       \\ \hline
7wii\_V          & 0.5918                 & 0.4209                  & 0.3573                   & \begin{tabular}[c]{@{}l@{}}GGACCGUCCGCCAACAACGCUCCCCGAAAGGGGAGCAGCG\\ GGAGGUCCA\end{tabular}                                                                                                                 \\ \hline
7xk1\_B          & 0.5556                 & 0.7460                  & 0.7842                   & \begin{tabular}[c]{@{}l@{}}CGGAGGUGGCGCAGUGGUAGCGCAGGCGAGUUCAACUCGC\\ CAGGCGCGGGUUCGAUUCCCGUCCUCCGGCCC\end{tabular}                                                                                          \\ \hline
8sh5\_R          & 0.5747                 & 0.2632                  & 0.3472                   & \begin{tabular}[c]{@{}l@{}}GCGAAACUGGCAGAAUCGGUUAUGAGUUAGUCGAGCGAGA\\ CACGCUCACCCACCUUUUUAGGUUGGCUAACCGUUCGCUC\\ GUUUUGA\end{tabular}                                                                        \\ \hline
8t2a\_R          & 0.5889                 & 0.3275                  & 0.4005                   & \begin{tabular}[c]{@{}l@{}}GGCUGCCGGAGUGCUUGUUGUCGUAGCCGGCAUGGAAAGA\\ CCAUGUGCUCGGCUACCCUUCGGGGUGUGAGCUACGGCAC\\ GACGGUGGUC\end{tabular}                                                                     \\ \hline
8fn2\_B          & 0.6964                 & 0.6011                  & 0.3474                   & \begin{tabular}[c]{@{}l@{}}GUCUGGUGGCCAUAGAAUCAAGGAACCACCUGAUCCCAUC\\ CCGAACUCAGAAGUUAAGCUUGAUAUCGGUGAUGAUAUUG\\ CGUUUUCGCGAGAAACUAGCGAACUGUCAGAA\end{tabular}                                               \\ \hline
8gxb\_B          & 0.6667                 & 0.1951                  & 0.1607                   & \begin{tabular}[c]{@{}l@{}}GAGCGUUGCUCGCAAGCGCCGCAUUGCACUUCGCGGCAGA\\ GGUGUUAAUAAAAAGAAGCG\end{tabular}                                                                                                      \\ \hline
8ine\_5          & 0.7417                 & 0.9157                  & 0.9529                   & \begin{tabular}[c]{@{}l@{}}GGGUACGGCCAUACUUCCCUGAAAACACCGAUUCCCCUCC\\ GAUCAUCGAAGUUAAGCAGGGACAGGCUUGGUUAGUACUC\\ GUGUCGGAGACGAACUGGGAACACCGAGUGCUGUACCCUU\end{tabular}                                       \\ \hline
8ipy\_8          & 0.5613                 & 0.2557                  & 0.5929                   & \begin{tabular}[c]{@{}l@{}}CAAUUCUCGACUCAGAAUAUUUGGCUUCCUCUUCGUUGAA\\ GAACGCAGCAAAAUGCGAUAAGCGAUAUGAGUUGCAAACA\\ UAAAAGAGUAUUAGGGGUUCGAACGCAAAGGCGCUCCCAG\\ UUGAAAUCUGGGAGUACAGCUCUUUCAGUCUCUUG\end{tabular} \\ \hline
\end{tabular}
}
\end{table*}

\end{appendices}

\bibliographystyle{abbrvnat}
\bibliography{appendix}